# A NEW COURSE ON CREATIVITY IN AN ENGINEERING PROGRAM: FOUNDATIONS AND ISSUES

Sophie Morin
Polytechnique Montréal
sophie.morin@polymtl.ca

Jean-Marc Robert
Polytechnique Montréal
jean-marc.robert@polymtl.ca

Liane Gabora
University of British Columbia
liane.gabora@ubc.ca

**ABSTRACT**
The importance of innovation in the world's economy, now undeniable, draws great attention to the need to improve organizations' creative potential. In the last 60 years, hundreds of books have been written on the subject and hundreds of webpages display information on how to be more creative and achieve innovation. Several North American and European universities offer graduated programs in creativity. However, building an effective and validated creativity training program is not without challenges. Because of the nature of their work, engineers are often asked to be innovative. Without aiming for a degree in creativity, could future engineers benefit from training programs in creativity? This article presents the conceptual framework and pedagogical elements of a new course in creativity for engineering students.

# 1. INTRODUCTION

"Creativity is just connecting things." – Steve Jobs (Wolf, 1996)

In this quote, Job's identifies a key element of the creative process (Wilkenfeld and Ward, 2001). In keeping with the trend in popular culture toward an increasingly demystified view of creativity, educators have started to consider the importance of developing their students' creativity.

However, engineering has been mostly perceived as a mathematical and technical discipline. Stouffer (2004) reports that in 1998, only 3% of the public surveyed (USA) thought "creative" and "engineers" were associated words. This is an unfortunate situation as engineers are constantly asked to be innovative.

This article presents a proposal for a new course dedicated to the development of cognitive abilities related to creativity in an engineering context. Based on a conceptual framework (CHC model), this course is expected to increase students' creative potential.

# 2. CONTEXT

The increased prominence of the word 'creativity' in everything from books and websites to consulting firms and university curricula, all speak to the perceived importance of creativity to the economy. In 2008, the United Nations published a report to explain and promote the value of a creative economy (CNUCED, 2008). In 2009 the European Union named the year "The European Year of Creativity and Innovation" (European, 2009).

To stay competitive, a company must constantlycome up with new and better ideas. Multidisciplinary teams are put together to study theorganizations' activities and find ways to improve the processes and products.

Typically, the fine arts (domains such as visualarts, music, dance, design, etc.) are most readily associated with the need for creativity. However, science could be considered equally creative (Sawyer, 2012). It is therefore important to help engineers and future scientists develop the competencies to optimize their creative potential.

## 2.1. ENGINEERING AND CREATIVITY

The need to study creativity and understand its underlying mechanisms is now globally accepted by the scientific community (Baillie and Walker, 1998, Richards, 1998, Sawyer, 2012). Numerous articles testify to the importance of the subject to the engineering community (Badran, 2007, Baillie and Walker, 1998, Chiu and Salustri, 2010, Cropley and Cropley, 2000, Richards, 1998). Being frequently in a position where there is a need for innovation, engineers should be able to actively participate in the creative process. Therefore, the development of creative skills as part of engineering training is essential, crucial, and indispensable. The next section will present a specific context in engineering education in which creativity training is progressing.

Polytechnique Montreal (PM) is a well-known engineering school in the world. With hundreds of graduates every year and a research budget of $84M, PM is a pillar in the engineering community.

PM's 13 baccalaureate programs were remodeled 10 years ago, including integrating project based learning (PBL) every academic year. PBL is associated with the integration of basic knowledge, problem analysis and solution, and instrumentally with the development or mobilization of creativity capacities. As such, students are primarily graded on the projects' results, leaving the creative process as an accessory. Creativity might be assessed in some cases but only as a partial characteristic of the solution.

In 2012, PM created a 12 hour workshop for doctorate students called "Creativity, yes we can". The course is offered every semester and is mandatory for all PhD students. Different educators offer different version of the workshop in English and in French. Both the content and pedagogical strategies result from the teacher's own research and experience.

PM acknowledges the need to develop its students' creativity, in order to produce better engineers.

### 2.1.1. The Potential Benefits

PM will have the opportunity to become a leader in the field and include creativity as an academic subject. This course will initially be offered as an option, especially interesting for students studying project management, technological innovation, ergonomics/Human factors, industrial engineering, etc. It could eventually become part of the curriculum of all engineering programs. The goal is to develop cognitive abilities linked to creative behaviours, every student could benefit from such training.

Looking even further and dreaming bigger, recurring creativity training in first and third year combined with an application in an integrative project, could completely transform how engineers and the public view the engineering profession.

Practice is required to develop and maintain new abilities (Badran, 2007, Cropley and Cropley, 2000, Sawyer, 2013). Cognitive abilities in mathematics and physics must continually be used and improved; it could be the same for creativity.

Learning the basics of how the creative process works and how it can be enhanced may improve and complete engineers' development. By better understanding the creative process and its cognitive implications, engineers are also expected to be more conversant with colleagues in other domains (design, marketing, etc.). Moreover, it is hoped that they will be better equipped to discuss, participate and come up themselves with new ideas to stimulate innovation.

### 2.1.2. The Challenges

Programs are already full with technical courses; there is not much room left for additional subjects. The implication of the Canadian Engineering Accreditation Board (CEAB) represents a big obstacle. Making room for a few hours of creativity training in the engineering curriculum will definitely be a big challenge.

To achieve this goal, many challenges must be overcome. First, pre-conceived ideas and judgments about creativity itself have to be addressed.

Students, educators, and managers, all have different perspectives on the subject. Managers will have to be convinced that creativity needs to be and can be developed. Educators will have to be adequately prepared to teach creativity and the cognitive abilities associated with it. Educators must also ensure that students are ready to develop their creativity. Finally, students will have to learn to be open minded about such a subject and its pedagogical implications even though the methods by which they develop competencies will differ from those of their other courses.

There are practical challenges to overcome. Working spaces can affect creativity

(Kahl, 2012). In a school, wooden desks, plastic chairs, white walls, and black boards are the norm, and these surroundings are not expected to bring out student creativity. If a complete creativity program existed, with multiple courses across engineering programs, it would be easier to justify the creation of a dedicated classroom organized to fit the needs of a creativity class.

## 2.2. THEORETICAL BACKGROUND

As Murdock (2011) explains, creativity does not have the same privileges as subjects like physics, in which elements are well defined and the pedagogical concerns are well established. Because creativity is still so connected to personal experiences and "recipe" books (Michalko, 2006, VanGundy, 2005) (even best sellers), many people consider they know enough to teach how to be more creative. Creativity can be taught in so many ways, we believe educators and most importantly students would benefit from more direction on how to create an effective creativity training program.

### 2.2.1. Intelligence and Cognitive Abilities

In the 1990's, J.B. Carroll develop an intelligence model that was well accepted by many scientists (Carroll, 1993, Newton and McGrew, 2010). He created the CHC model (Cattel-Horn-Carroll) to classify various cognitive abilities. Some of these specific abilities can be related to the creative process and become the foundation of a training program.

Three broad CHC abilities are particularly meaningful: fluid reasoning (Gf), comprehensionknowledge (Gc) and long-term storage and retrieval (Glr). Gf consists of the abilities related to executive processes, i.e., the different aspects of reasoning. Gc represents rather the accumulation of knowledge. It used to be referred to as crystallized intelligence. Thirdly, Glr is closely linked to abilities that foster creativity. The 13 narrow abilities included in this broad ability can all be associated with different phases of a creative behaviour (e.g., associative memory, meaningful memory, ideational fluency, naming facility, etc.).

To adapt the CHC model to creativity, we suggest creating a second level of abilities to further classify the Glr narrow abilities. With Gabora's research on creativity on a cognitive level (Gabora, 2010, Gabora and Saab, 2011), we added specific intermediate cognitive abilities: encoding, potentiality, divergent thinking, and convergent thinking. We believe that building pedagogical content and teaching strategies aimed at developing those specific abilities may improve students' creative potential.

The cognitive ability we call "encoding" refers to the way we store information in our brains. Neurons form a web-like network structure (Hopfield, 1982). Because creativity relies on the capacity to make remote associations between known elements, we hypothesize that individuals with wider and richer neurons networks will have greater creative potential. If we can help learners develop richer cognitive networks, we could help them improve their ability to draw connections.

It has been shown that creative tasks ranging from highly constrained analogy problem solving to highly unconstrained art making involve the "actualization of potentiality" (Gabora and Saab, 2011). By this we mean the creative task induces a cognitive state that is an amalgam of potentially disparate thoughts, memories, and concepts, which feels subjectively like a 'half-baked idea' and which can be mathematically described as a state of superposition (Aerts et al., 2013, Gabora and Aerts, 2002, 2009). It can unfold different ways depending on the different contexts it interacts with, i.e., perspectives it is considered from. We suggest that by teaching students to be comfortable with half-baked ideas, and trust in their own abilities to develop them, students' creative potential could be enhanced.

Lastly, there is broad consensus that divergent thinking and convergent thinking, are at the center of the creative process. A creativity training program should cultivate these specific kinds of thinking processes.

**2.2.2. Pedagogical Elements**
Research by Scott and his colleagues (Scott et al., 2004a, b) shows the latitude taken in every creativity training program aspect (course design, techniques, media, etc.). They conducted a meta-analysis of 70 different creativity trainings. Though 10 years old, these articles contain the most recent and detailed information available on training programs. Their four conclusions are presented in the Table 1 (Scott et al., 2004a, b).

Table 1 – Scott's Conclusions

| **Conclusions** |
| --- |
| • Training based on valid conception of the cognitive activities underlying creative efforts. |
| • Lengthy, relatively challenging with various discrete cognitive skills, and associated heuristics. |
| • Articulation of principles should be followed by illustrations of their application using material based on "real-world" cases or other contextual approaches. |
| • Presentation should be followed by a series of exercises appropriate to the domain at hand. |

These conclusions are very general and lack details (e.g., how long is lengthy, what cognitive activities or skills should be developed, etc.). However, certain elements can be taken into account when creating a creativity training program.

Finally, we suggest that activities or questions asked in creativity measuring tests (Purdue Creativity Test, Finke's Creative Invention Task, etc.) be directly and explicitly used to train the brain to think creatively. These tests often incorporate tasks we are not accustomed to that could be put to good use. For example, in every class we could choose a different object (not the usual brick) and practice the "alternate uses" exercise (famous test created by Guilford to assess creativity). Iterative practice of this task may help students develop the cognitive abilities linked to this task. When a specific project will require the use of a similar exercise, the learnings may be transferable. A great number of ideas can be found in those tests to build an effective training program.

**2.3. THE NEW TRAINING PROGRAM**
The course will be based on the hypothesis that developing cognitive abilities linked to creativity may enhance creative potential (encoding, potentiality, divergent/convergent thinking). Numerous and various pedagogical strategies will be used to achieve that goal (serious games, creativity approaches, conferences, individual and group projects, etc.). The next section will present the content as well as some of the chosen strategies with more details.

### 2.3.1. The Content
The course's program will include several different aspects of creativity, some more traditional than others. Theoretical elements will have to be presented and discussed but the development of cognitive abilities will be the center of attention.

First of all, influential factors (environmental, social, personal, organizational, etc.) will be discussed. Their impact on creativity is well documented, and understanding them is important for metacognitive processes (Barack and Goffer, 2002, Kowaltowski et al., 2010, Richards, 1998). Some educators base their training program on those factors. We agree with the importance of the knowledge of these factors but we also believe it is only a portion of the training to improve creative potential.

Also, a few creativity approaches (methods, techniques, activities) will be explained and experienced (e.g.: Mind-mapping, six hats, SCAMPER, bionics). So many methods and techniques exist, we think it is more profitable for students to explore a limited number of them and have the opportunity to try them than simply to describe globally dozens of those approaches. A hands-on experience is expected to be more efficient in the development of cognitive abilities than a masterful presentation.

Creativity involves domain-general abilities as well as domain specific abilities (Sawyer, 2012). The course will include both. Engineering examples and projects will be used to transfer general cognitive abilities to engineering specific domains. The multiple integrative projects already in the curriculum will provide another opportunity to use the general abilities developed.

Many researchers have described the importance of "lateral" or "flat" knowledge in the creativity process (Litzinger et al., 2011, Yeh, 2011). Being an expert in one subject is not the best path to creative behaviour. It is important to develop a deep understanding of a certain subject but it is also essential to learn many things in other domains. If one wants to connect remote elements in order to produce creative ideas, he has to know a lot of different topics (Bonnardel, 2000). To take into account these findings, the course will include artistic projects (writing, singing, dancing, drawing, etc.) and conferences in different themes like marketing, improvisation and invention. That will allow students to expand their knowledge hierarchy and stimulate remote associations.

Finally, the element driving this new course, the development of cognitive abilities, will be a major part of the content. In every class, exercises will be devoted to developing abilities in encoding information, resolving an analogy and associative and analytical thinking. Simple activities, often in the form of serious games, will allow students to better understand how they think and what other mechanisms they could use to think "differently". Not necessarily in a specific context, these serious games will help learners develop cognitive abilities without focusing them on a specific problem to resolve. Because there is evidence that improving creativity in a subject does not necessarily help in another (Baer and Kaufman, 2005), we hope to develop general abilities that readily transfer to other domains.

### 2.3.2. The Pedagogical and Practical Aspects
Various pedagogical strategies will be used in this new course. Conferences, discussions,

serious games, individual and group projects, etc., will allow students to look at creativity in different ways. The Table 2 below presents some examples of pedagogical choices. The most unusual activity is probably the warm-up exercises. Based on the need to practice to develop skills (Rasmussen, 1983), i.e. repeat the same movements many times in order to master them, these serious games sessions will allow learners to become more familiar with the cognitive abilities linked with creativity (Cropley and Cropley, 2000).

Table 2 – Pedagogical Choices

| Pedagogical Activities | Subject/Objective |
| --- | --- |
| Warm-up exercises<br>• 1/2 h at the beginning of every class, different versions of the same exercises, individual and group | Cognitive abilities (associative and analytical thinking, encoding, potentiality) |
| Conferences<br>• 1 1/2 h, invited speaker, presentation + exercises | Marketing, Improvisation, Invention |
| Log book<br>• Individual task, spread on all semester, evaluate a few times during the semester, students note observations, ideas, problems, solutions, etc. | Being a better observer, more critical towards what you see |
| Artistic Project<br>• Individual project, students choose their subject, first half of semester, presentation to group | Expand knowledge in different domains |
| Engineering Project<br>• Group project, students choose the problem they want to tackle, second half of semester, presentation to group | Apply new cognitive abilities |

The room is to become a practical matter. A room very different from typical classrooms will be necessary to take students out of their usual surroundings and comfort zone. Chairs, tables, wall colors, posters, lights, decoration, etc., everything will be out of the ordinary. Ideally, each semester, students could create a new environment by bringing personal pieces to the space. We could take advantage of this setting to create different exercises every semester.

     Assessing creativity has been the source of numerous articles (Charyton et al., 2011, Clary et al., 2011, Shah et al., 2003, Villalba, 2012). There is no consensus as to how creativity should be evaluated in a specific context. For assessing creativity in an educational context, we favor the enhancement of the creative potential instead of the level of creativity. In short, we want to assess the progress made by students and their efforts towards these improvements and not only the actual creative performance. To achieve that goal, we present in Table 3 a few examples of evaluation choices.

     The CEDA instrument has been developed by Charyton (Charyton, 2014, Charyton et al., 2008, Charyton et al., 2011). Especially made for engineering students, it

gives a score based on a judging process. By giving the test on the first and the last class, we expect to find an improvement in the individual scores. To minimize any doubts around the evaluation process of this new course, other evaluation methods will assess different aspects of the creative process.

Table 3 – Evaluation Choices

| Evaluation choices | Aspects |
| --- | --- |
| • CEDA (Creative Engineering Design Assessment) | • Creative potential improvements |
| • Participation (in class) | • Personal implication (intrinsic motivation) |
| • Oral presentation (creativity approaches) | • Research, originality, convincing abilities |
| • Log book | • Observation ability, Openness to problem finding |
| • Artistic project/Engineering project | • Individual and group work |
| • Mid-term (mind map) | • Knowledge of the subject (theories, models, methods, etc.) |

## 4. CONCLUSIONS

We presented a new course that will allow the engineering community to see how creativity can be taught and what impacts such a course may have on future engineers. This could be the beginning of a new vision of creativity training programs for engineers and other scientists. A number of domains (e.g., medicine or business) could take advantage of these learnings and integrate creativity into their program. Fostering innovation by enhancing creative cognitive abilities could be profitable in numerous subjects.

Additionally, organizations could definitely benefit from these research and results. They could include creativity training programs as part of their continuous training program and offer employees recurring creativity workouts to develop and maintain their creative abilities. We hope that innovative engineering feats will emerge from these next steps in creativity training research.